\documentclass[aps,prl,twocolumn,preprintnumbers,groupedaddress]{revtex4}

\usepackage{amssymb}
\usepackage{epsfig} 
\usepackage{amsmath,color} 
\usepackage{graphicx} 
\usepackage{ulem} 
\usepackage{tabularx}
\usepackage{slashed}
\usepackage{hyperref}

\usepackage{tikz}
\usetikzlibrary{positioning,arrows}
\usetikzlibrary{decorations.pathmorphing}
\usetikzlibrary{decorations.markings}

\newcommand{\beq}{\begin{equation}}
\newcommand{\eeq}{\end{equation}}
\newcommand{\bea}{\begin{eqnarray}}
\newcommand{\eea}{\end{eqnarray}}
\newcommand{\ba}{\begin{array}}
\newcommand{\ea}{\end{array}}
\newcommand{\bi}{\begin{itemize}}
\newcommand{\ei}{\end{itemize}}
\newcommand{\bn}{\begin{enumerate}}
\newcommand{\en}{\end{enumerate}}
\newcommand{\bc}{\begin{center}}
\newcommand{\ec}{\end{center}}
\renewcommand{\l}{\left}
\renewcommand{\r}{\right}

\newcommand{\eq}[1]{Eq.~(\ref{#1})}
\newcommand{\eqs}[2]{Eqs.~(\ref{#1}) and (\ref{#2})}

\newcommand{\eV}{\mathinner{\mathrm{eV}}}

\begin{document}

\preprint{FTUV-16-08-05}
\preprint{IFIC-16-60}

\title{Resurrection of large lepton number asymmetries from neutrino flavor oscillations}


\author{Gabriela Barenboim$^1$}
\email[]{Gabriela.Barenboim@uv.es}
\author{William H.\ Kinney$^2$} 
\email[]{whkinney@buffalo.edu}
\author{Wan-Il Park$^1$}
\email[]{Wanil.Park@uv.es}
\affiliation{$^1$
Departament de F\'isica Te\`orica and IFIC, Universitat de Val\`encia-CSIC, E-46100, Burjassot, Spain}

\affiliation{$^2$ Dept. of Physics, University at Buffalo, 239 Fronczak Hall, Buffalo, NY 14260-1500}


\date{\today}

\begin{abstract}
We numerically solve the evolution equations of neutrino three-flavor density matrices, and show that, even if neutrino oscillations mix neutrino flavors, large lepton number asymmetries are still allowed in certain limits by Big Bang Nucleosynthesis (BBN). 
\end{abstract}


\maketitle


\section{Introduction}

Despite the fact that the baryon number asymmetry of the Universe is constrained to be $B \sim 10^{-10}$ by Big Bang nucleosynthesis (BBN) \cite{Agashe:2014kda} and observations of the Cosmic Microwave Background (CMB) \cite{Ade:2015xua}, the universe is allowed to have a large lepton number asymmetry ($L$ defined similarly to $B$ but for leptons) as long as such an asymmetry is associated with neutral particles.
In particular, a large asymmetry of neutrinos (e.g. $L \gtrsim \mathcal{O}(1))$ is quite attractive in view of its impacts on cosmology (for example, as a solution to the problem of topological defects via a symmetry non-restoration \cite{Riotto:1997tf,Bajc:1997ky}, generation of $B$ from $L$ via sphaleron \cite{Liu:1993am}, 
and/or as contribution to the extra relativistic species $\Delta N_{\rm eff}$ which may lead to a better fit to cosmological data).  
Even if the total lepton number vanishes, $L=0$, the asymmetry $L_\alpha$ (for a neutrino flavor $\nu_\alpha$) could be large enough to have an impact on the generation of $B$ \cite{MarchRussell:1999ig} and $\Delta N_{\rm eff}$.

The main constraints on large neutrino asymmetries come from BBN (especially the abundance of $^4He$) \cite{Iocco:2008va,Cyburt:2015mya} and $\Delta N_{\rm eff}$, which is constrained by both BBN and CMB observations \cite{Ade:2015xua}.
In particular, BBN strongly constrains the asymmetry of electron neutrinos such that the degeneracy parameter ($\xi_\alpha \equiv \mu_\alpha/T$, with $\mu_\alpha$ being the chemical potential of $\nu_\alpha$ and and $T$ the temperature) is constrained as \cite{Mangano:2011ip}
\beq
-0.018 \leq \xi_e \leq 0.008 \Rightarrow -4.5 \lesssim 10^3 L_e \lesssim 2.0,
\eeq
while recent Planck satellite data of CMB observations constrain $\Delta N_{\rm eff} \lesssim 0.36$ at $95$ \% CL (Planck, TT,TE,EE+lowP+BAO) \cite{Ade:2015xua} which conventionally translates to
\beq \label{xi-Planck-con}
|\xi_{\mu, \tau}| \lesssim 0.89 \Rightarrow |L_{\mu,\tau}| \lesssim 0.24
\eeq
where $L_\alpha \equiv (n_\alpha - n_{\bar{\alpha}})/n_\gamma$ with $n_\alpha$/$n_{\bar{\alpha}}$ and $n_\gamma$ being the number density of $\nu_\alpha$/$\bar{\nu}_\alpha$ and photons, respectively.

Meanwhile, there has been a series of works showing that neutrino oscillations in the early universe mix three neutrinos such that any asymmetry $L_{\mu,\tau}$ which is established well before BBN is converted significantly to $L_e$ \cite{Lunardini:2000fy,Dolgov:2002ab,Wong:2002fa,Abazajian:2002qx,Mangano:2010ei,Mangano:2011ip,Castorina:2012md}.
As a result, BBN requires $|L_{\mu,\tau}| \lesssim 0.1$ which translates to $\Delta N_{\rm eff} \lesssim 0.07$ \cite{Mangano:2011ip,Castorina:2012md}.  
However, in these numerical simulations, the quantum kinetic equations of neutrino/anti-neutrino density matrices were solved using a scheme such that the mixed three-flavor neutrino system was handled as successive effective two-flavor systems ($\nu_\mu$-$\nu_\tau$ and $\nu_e$-$\nu_{\mu,\tau}$) before and after $\nu_\mu$-$\nu_\tau$ equalization, as can be seen by the fact that the $\nu_\mu$-$\nu_\tau$ degeneracy once established is never lifted, which is not consistent with the three mixing angles being non-zero, as will be discussed below. 
For the evolution up to the point of $\nu_\mu$-$\nu_\tau$ equilibrium, the two-flavor description is enough since $\nu_e$ participates in the oscillations only afterwards.
However, once $\nu_e$ is involved, the evolution of the mixed three-flavor system becomes quite complicated, and the $\nu_\mu$-$\nu_\tau$ equalization may not be maintained any more. 
In addition, the choice of neutrino mixing parameters has significant impact on the final asymmetry of each flavor after the mixing and oscillation effects settle the system to an equilibrium state.
More importantly, the total asymmetry $L$ does not need to be small as long as its contribution to $B$ is suppressed by symmetry non-restoration. 
In this regard, it is worth while to re-examine the impact of \textit{three-flavor oscillations} of neutrinos on the BBN bound of the lepton number asymmetries. 

In this letter, we argue that effective two-flavor description of the mixed three-flavor neutrino system does not necessarily capture the real physics of neutrino oscillations in the early universe. We demonstrate our argument by presenting the numerical solution to the three-flavor evolution equations, which is different from the results in earlier work based on a two-flavor effective description.
We also show that BBN still allows large asymmetries which can lead to $\Delta N_{\rm eff} \sim \mathcal{O}(1)$.

\section{Two- or three-flavor description?}

The masses and mixing parameters of neutrino oscillations are now measured to be \cite{Agashe:2014kda,NOvA}
\bea
\Delta m_{21}^2 &=& 7.53^{+0.18}_{-0.18} \times 10^{-5} \eV^2 
\\
\Delta m_{31}^2 &\simeq& \Delta m^2_{32} =  2.67 \pm 0.12 \times 10^{-3} \eV^2
\eea
and 
\bea
\sin^2 2\theta_{12} &=& 0.846 \pm 0.021
\\
\sin^2 2\theta_{13} &=& 0.093 \pm 0.008    
\\
\sin^2 \theta_{13} &=& 0.40^{+0.03}_{-0.02} \ (0.63^{+0.02}_{-0.03}) 
\eea
where $\theta_{ij}$ are the mixing angles in the Pontecorvo-Maki-Nakagawa-Sakata (PMNS) mixing matrix \cite{Maki:1962mu,Pontecorvo:1957cp} whose CP-violating phase is set zero here.
In the early universe, the oscillations of neutrino flavors can be described by the evolutions of neutrino/anti-neutrino density matrices. 
For a mode of momentum $p$, the density matrices in the flavor basis of three active neutrinos $(\nu_e, \nu_\mu, \nu_\tau)$ can be expressed in terms of polarization vectors $\mathbf{P}/\bar{\mathbf{P}}$ and Gell-Mann matrices $\lambda_i \ (i=1-8)$ as
\beq
\rho_p = \frac{1}{3} \sum_{i=0}^8 P_i \lambda_i, \quad \bar{\rho}_p = \frac{1}{3} \sum_{i=0}^8 \bar{P}_i \lambda_i
\eeq   
where $\lambda_0$ is the $3\times3$ identity matrix.
Then, the evolution equations of $\rho_p$ and $\bar{\rho}_p$ are given by \cite{Sigl:1992fn,Pantaleone:1992eq} (see also \cite{Blaschke:2016xxt})
\bea 
\label{eom-rho}
i \frac{d\rho_p}{dt} &=& \l[ \Omega + \sqrt{2} G_F \l( \rho - \bar{\rho} \r), \rho_p \r] + C \l[ \rho_p \r] \\
\label{eom-rhobar}
i \frac{d \bar{\rho}_p}{dt} &=& \l[ -\Omega + \sqrt{2} G_F \l( \rho - \bar{\rho} \r), \rho_p \r] + C \l[ \bar{\rho}_p \r]
\eea 
In the above equations,
\beq \label{Omega}
\Omega = \frac{M^2}{2 p} - \frac{8 \sqrt{2} G_F p E_\ell}{3 m_W^2}
\eeq
where $M^2$ is the mass-square matrix of neutrinos in flavor-basis, $G_F$ the Fermi constant, $m_W$ the mass of $W$-boson, $E_\ell = {\rm diag}(E_{ee}+E_{\mu\mu}, E_{\mu\mu},0)$ the energy density of charged leptons, $\rho = (1/2 \pi^2) \int_0^\infty \rho_p p^2 dp$ (and similarly for $\bar{\rho}$), and $C[\dots]$ is the collision term.

Practically, we numerically solve the equations of motion (EOM) of $P_i$ and $\bar{P}_i$ derived from \eqs{eom-rho}{eom-rhobar}.
Those equations are mixed in a complicated way, and it is non-trivial to get an insight of what may happen unless a numerical integration is performed.
It is also difficult to see if the maintenance of $\nu_\mu$-$\nu_\tau$ equalization in an effective two-flavor description taken in earlier works still is valid in this case.
However, it is instructive to note that, when one of the mixing angles is set zero with $\theta_{23}=\pi/4$, the mass-square matrix $M^2$ has a special pattern (for example, if $\theta_{13}=0$, then $M^2_{12} = - M^2_{13}$ and $M^2_{22} = M^2_{33}$).
In this case, ignoring the subdominant collision terms in \eqs{eom-rho}{eom-rhobar}, one can see that some pairs of $P_i^\pm$s (for example, $P_1^-$-$P_4^-$ and $P_2^+$-$P_5^+$ where $P_i^\pm \equiv P_i \pm \bar{P}_i$) are likely to be driven in exactly opposite way.  
As a result, it becomes possible to have $P_3^- - \sqrt{3} P_8^- = 0$ and $d \l( P_3^- - \sqrt{3} P_8^- \r)/ dt = 0$ simultaneously, and this implies that, once $\nu_\mu$-$\nu_\tau$ equalization is achieved, it is likely to be maintained even if another non-zero mixing becomes active.
This is the case in which the two-flavor description can be applicable.
However, if all the mixing angles are non-zero as the accumulated neutrino oscillation data indicate, or $\theta_{23} \neq \pi/4$ even if $\theta_{12}=0$ or $\theta_{13}=0$, the special pattern of the square-mass matrix disappears, and there is no reason to expect $\nu_\mu$-$\nu_\tau$ equalization to be maintained once the second and/or third mixing get involved. 
Hence, we can expect that the two-flavor description may be applicable only to that limited case, which does not apply in view of the current observational data in neutrino oscillation experiments. 
In the next section, we will show that this is in fact the case.

\section{Results of three-flavor numerical integration}
In our numerical analysis, $M^2$ was taken to correspond to a normal hierarchy of neutrino masses.
Also, since a precise treatment of collision terms has only minor effect in the scope of this letter (see for example \cite{deSalas:2016ztq}), 
we take for simplicity $C[\rho_p] = - i D_{\alpha \beta} [\rho_p]_{\alpha \beta}$ for $\alpha \neq \beta$ only, and similarly for $C[\bar{\rho}_p]$ \cite{Dolgov:2002ab}.
The initial condition for the simulations was set as
\beq \label{rho-p-initial}
\rho_p = f(y,0)^{-1} {\rm diag}(f(y,\xi_e), f(y,\xi_\mu), f(y,\xi_\tau)), 
\eeq
and similarly for $\bar{\rho}_p$ but with $\xi_\alpha \to -\xi_\alpha$, where $f(y,\xi_\alpha)=\l( e^{y-\xi_\alpha}+1 \r)^{-1}$ is the occupation number of $\nu_\alpha$ for a mode $y \equiv p/T$.

In the presence of charged lepton backgrounds and/or collisional dampings, the dynamics of the occupation number of a mode is not oscillation-like, but transition-like.
In this case, the dynamics of flavor asymmetries (as a mode-integrated collective behavior) can be mimicked by a typical mode ({\it i.e.}, corresponding to the averaged momentum or close to it) even without the self interaction term ({\it i.e.}, the term proportional to $\rho-\bar{\rho}$ in \eq{eom-rho} or (\ref{eom-rhobar})), modulo an overall normalization \cite{Abazajian:2002qx}.
We take this single mode approach with $y=3.15$ which is nearly the same as the mode of average-momentum, but in order not to miss specific phenomena caused by self-interaction (e.g., blocking of transition \cite{Dolgov:2002ab}), we keep the self-interaction term in a way that $\rho^-$ ($\equiv \rho-\bar{\rho}$) is replaced by $\rho_p^-$ ($\equiv \rho_p-\bar{\rho}_p$), normalized initially to match $\rho^-$. 
In order to see the result in terms of the lepton number asymmetries, the initial occupation numbers of our reference mode were normalized to match the initial lepton number asymmetries accounting all modes.  

\begin{figure}[h]
\begin{center}
\includegraphics[width=0.5\textwidth]{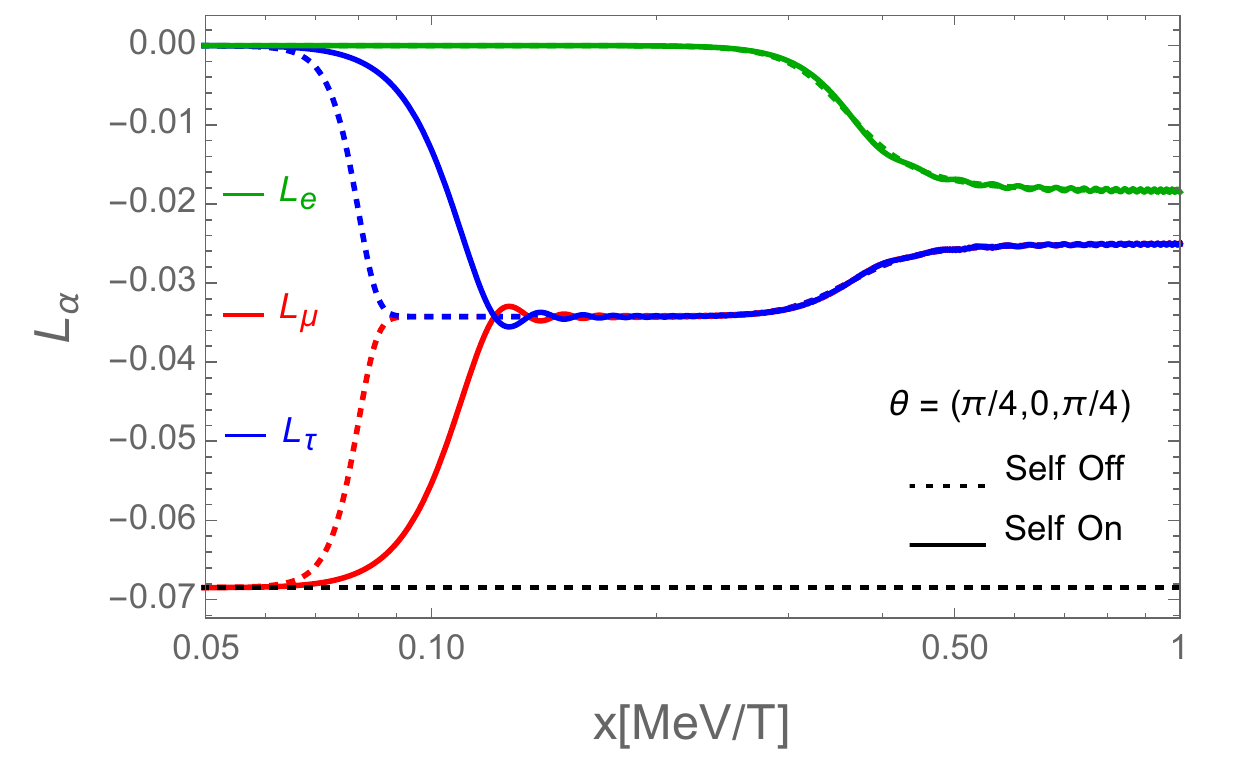}
\caption{Evolutions of $L_\alpha$ for $\theta=(\theta_{12},\theta_{13},\theta_{23})$ and $(\xi_e,\xi_\mu,\xi_\tau) = (0,-0.1,0)$ with self-interaction switched on/off (solid/dotted lines). 
Green/red/blue line is for $L_e/L_\mu/L_\tau$.
Black dotted line is the total asymmetry.
}
\label{fig:validity}
\end{center}
\end{figure}

The validity of our approach was checked by reproducing some results in earlier works, for example, as shown in Fig.~\ref{fig:validity} (see Fig.~5 of Ref.~\cite{Dolgov:2002ab} for comparison).
The figure shows that the main features of the evolution of $L_\alpha$ governed by \eqs{eom-rho}{eom-rhobar} are captured by our simplified approach, proving the validity of our approach. 
The minor difference of the amplitude of synchronized oscillation (which depend on $|\eta_\alpha|$ or $\xi_\alpha$) may be the difference between effective two-flavor description and three-flavor full description.
If the initial asymmetries are large enough and are not forced to obey a specific pattern (e.g., equal and opposite), the evolution of the asymmetries appears to be essentially independent of the self-interaction term.
This implies that for aligned initial asymmetries, when the self-interaction is large enough, $\mathbf{P}^+$ hardly deviates from the direction of $\mathbf{I}^- \equiv \sqrt{2} G_F \int \mathbf{P}^- p^2 dp/(2 \pi^2)$ (or simply $\mathbf{P}^-$ in our simplified simulation). 

\begin{figure}[t]
\begin{center}
\includegraphics[width=0.475\textwidth]{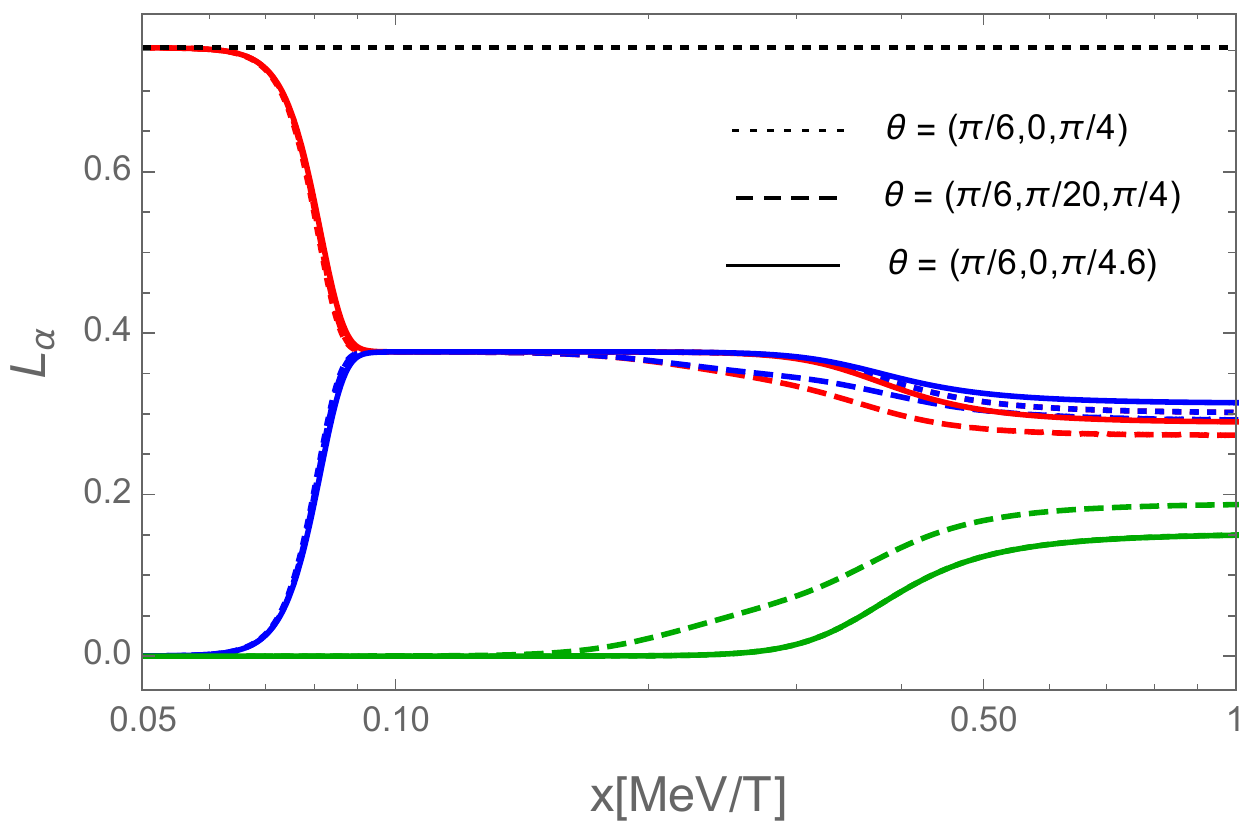}
\caption{Evolutions of $L_\alpha$ for $\theta=(\theta_{12},\theta_{13},\theta_{23})$ and $(\xi_e,\xi_\mu,\xi_\tau) = (0,1,0)$ with self-interaction switched on.
Color scheme is the same as Fig.~\ref{fig:validity}.
}
\label{fig:e-mu-tau-010}
\end{center}
\end{figure}

As our first new result, Fig.~\ref{fig:e-mu-tau-010} shows the evolution of $L_\alpha$ with different sets of mixing angles.
The self-interaction term did not make any meaningful change in this case, as expected.
As shown in the figure, the first dynamics takes place due to $\theta_{23} \sim \pi/4$ which mixes $\nu_\mu$ and $\nu_\tau$ completely, leading to $L_\mu=L_\tau$ irrespective of the value of $\theta_{23}$ due to frequent collisions.
At later time, collisions become inefficient.
In this circumstance, if $\theta_{23}=\pi/4$ and $\theta_{13}=0$ (dotted lines), this equalization is maintained even if non-zero $\theta_{12}$ gets involved later.
%
\begin{figure}[h]
\begin{center}
\includegraphics[width=0.5\textwidth]{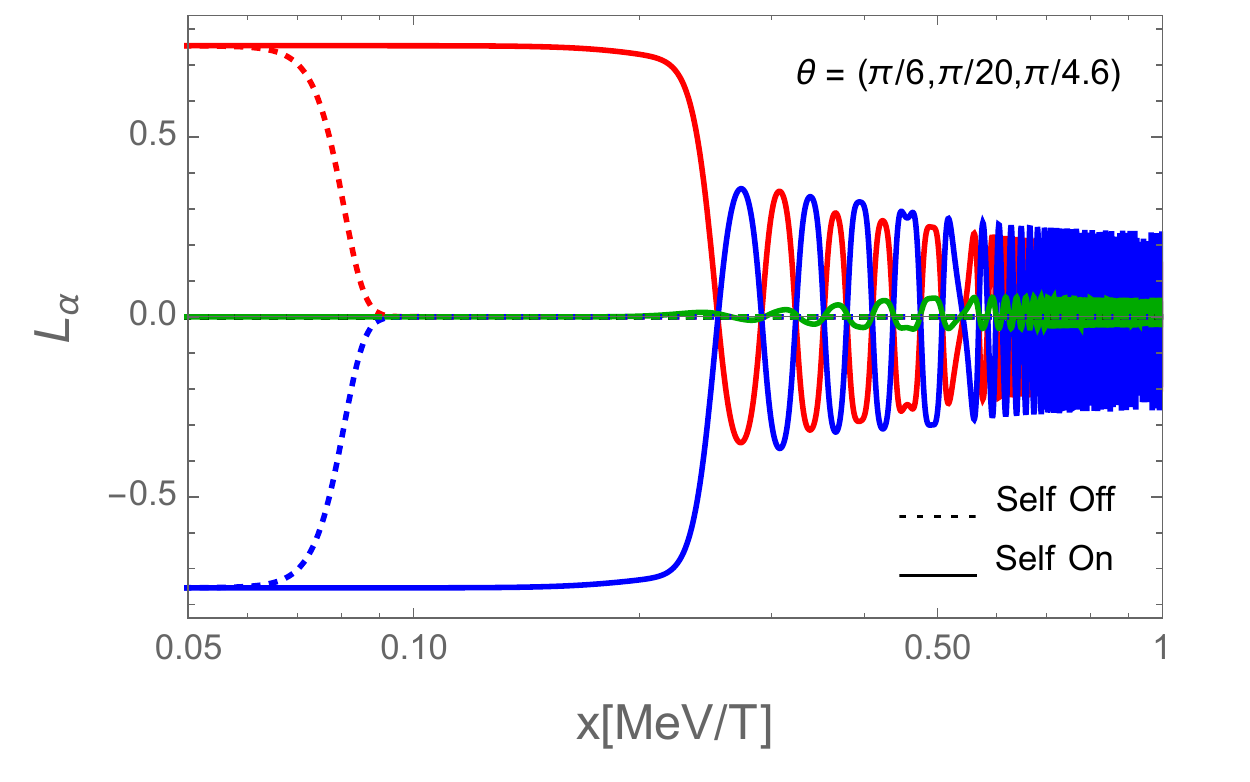}
\caption{Evolutions of $L_\alpha$ for $\theta=(\theta_{12},\theta_{13},\theta_{23})$ and $(\xi_e,\xi_\mu,\xi_\tau) = (0,1,-1)$.
Color scheme is the same as Fig.~\ref{fig:validity}.
}
\label{fig:0net-single}
\end{center}
\end{figure}
\begin{figure}[h]
\begin{center}
\includegraphics[width=0.5\textwidth]{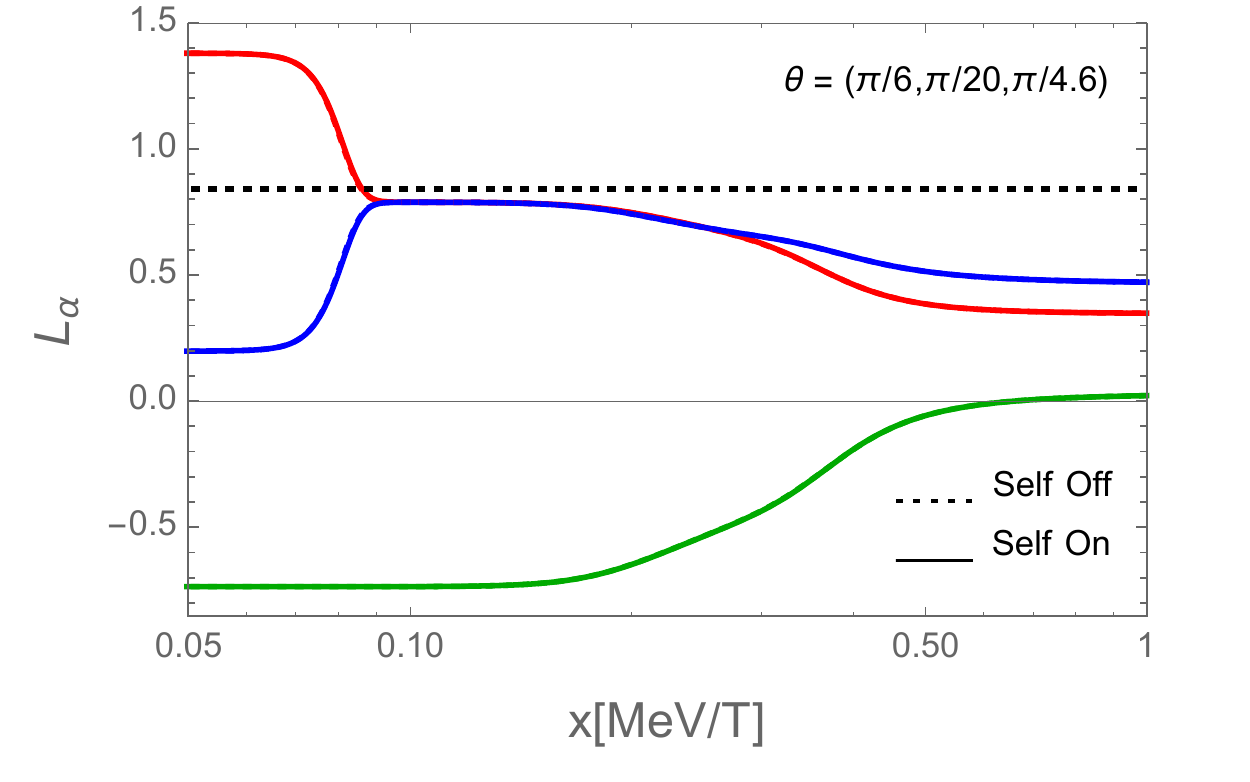}
\caption{Evolutions of $L_\alpha$ for $\theta=(\theta_{12},\theta_{13},\theta_{23})$ and $(\xi_e,\xi_\mu,\xi_\tau) = (-1.0,1.6,0.3)$.
Color scheme is the same as Fig.~\ref{fig:validity}.
Dotted lines (the case of ``Self Of'') were overlapped by solid lines.
}
\label{fig:net-single}
\end{center}
\end{figure}
Checking the dynamics of all components of polarization vectors, we found that the reason for such a behavior is exactly what discussed in the previous section.
The same behavior appears if $\theta_{12}$ is set zero instead of $\theta_{13}$.
On the other hand, if all the mixing angles are non-zero (dashed lines) or $\theta_{23} \neq \pi/4$ (solid lines), the equalization is broken, as the second dynamics appears due to another mixing.
Therefore, we conclude that, for the neutrino mixing parameters measured so far, $L_\mu \neq L_\tau$ as a result of neutrino mixings.

Obviously, the final $L_\alpha$ depends on $L$.
So, we now consider some initial values of $L_\alpha$ for $L=0$ and $L\neq0$ cases as shown in Fig.~\ref{fig:0net-single} and Fig.~\ref{fig:net-single}, respectively.
In the case of Fig.~\ref{fig:0net-single}, due to the equal and opposite asymmetries of $\nu_\mu$ and $\nu_\tau$, neutrino self-interaction blocks the $\nu_\mu$-$\nu_\tau$ mixing until the dynamics due to the non-zero $\theta_{13}$ becomes active.
This phenomenon was observed already in an earlier work \cite{Dolgov:2002ab}, but the subsequent synchronized oscillation was not clear in the result, in contrast to our case.
The large synchronized oscillation seems to be due to the delayed mixing of $\nu_\mu$-$\nu_\tau$ that is dominated by the vacuum contribution in \eq{Omega}.
The final asymmetries depend on the mixing angles and configuration of $L_{\alpha,0}$.
However, for $L=0$, even if $L_{e,0}=0$, the oscillation-averaged values turn out to be
\beq
|L_e| \sim |L_{\mu,\tau}| \lesssim 10^{-2} |L_{\mu,0}|
\eeq
where $L_{\alpha,0}$ is the initial asymmetry of $\nu_\alpha$, and $\xi_{\alpha,0} \lesssim 1$ was assumed.
Hence, in this case we end up the same conclusion as earlier works.

Contrary to the case of $L=0$, if $L\neq0$, one can take arbitrary initial values of $L_\alpha$.
This means that, as shown in Fig.~\ref{fig:net-single}, at the late time equilibrium it is possible to have small $L_e$ but large $|L_{\mu,\tau}|$ which can result in $\Delta N_{\rm eff} \sim \mathcal{O}(1)$.
Note that the net asymmetry $L$ can be large enough to suppress the conversion of $L$ to $B$ by symmetry non-restoration \cite{Bajc:1999he}.
This is our main result.

\section{Conclusions}

In this letter, we showed that, contrary to the widely held conventional expectation, lepton number asymmetries of neutrinos can be quite large while keeping the asymmetry of electron-neutrino small enough to satisfy the BBN bound. Large asymmetries of muon- and tau-neutrinos are expected to be constrained mainly by CMB through $\Delta N_{\rm eff}$ (extra neutrino species or ``dark'' radiation), but the asymmetries are better constrained in terms of neutrino mass-eigenstates instead of flavor-eigenstates, as will be discussed in a forthcoming paper \cite{newCMBcons}.

In earlier works in the literature, an effective two-flavor description after the first transition between muon- and tau-neutrinos was used by fixing the asymmetries of $\nu_\mu$ and $\nu_\tau$ equal.
However, such a setting is questionable in the presence of \textit{three non-zero} mixing angles. 
In addition, neither BBN nor CMB data forbid a large non-zero total lepton number asymmetry.
Motivated by these observations, we numerically integrated the quantum kinetic equations of the full three-flavor density matrices of neutrinos and anti-neutrinos, and found that the asymmetries of $\nu_\mu$ and $\nu_\tau$ after all the transitions are finished before BBN are actually different, and can be large enough to result in $\Delta N_{\rm eff} \sim \mathcal{O}(0.1-1)$ which may lead to a better fit to cosmological data \cite{newCMBcons}.

For large arbitrary initial lepton number asymmetries well before BBN, the stringent BBN bound on the asymmetry of electron-neutrinos appears to require a fine tuning of the initial condition.
However, such a tuning is certainly allowed by data, and could well be explained by some physics beyond the standard model.

\section{Acknowledgements}
The authors are grateful to Sergio Pastor for his helpful comments. 
GB and WIP also acknowledge support from the MEC and FEDER (EC) Grants SEV-2014-0398 and FPA2014-54459 and the Generalitat Valenciana under grant PROMETEOII/2013/017. This project has received funding from the European Union's Horizon 2020
research and innovation programme under the Marie Sklodowska-Curie grant
Elusives ITN agreement No 674896  and InvisiblesPlus RISE, agreement No 690575. 
WHK is supported by the U.S. National Science Foundation under grant NSF-PHY-1417317. WHK thanks the University of Valencia, where part of this work was completed, for hospitality and support.



\begin{thebibliography}{99}


\bibitem{Agashe:2014kda} 
  K.~A.~Olive {\it et al.} [Particle Data Group Collaboration],
  Chin.\ Phys.\ C {\bf 38}, 090001 (2014).
  doi:10.1088/1674-1137/38/9/090001


\bibitem{Ade:2015xua} 
  P.~A.~R.~Ade {\it et al.} [Planck Collaboration],
  arXiv:1502.01589 [astro-ph.CO].

\bibitem{Riotto:1997tf} 
  A.~Riotto and G.~Senjanovic,
  Phys.\ Rev.\ Lett.\  {\bf 79}, 349 (1997)
  doi:10.1103/PhysRevLett.79.349
  [hep-ph/9702319].


\bibitem{Bajc:1997ky} 
  B.~Bajc, A.~Riotto and G.~Senjanovic,
  Phys.\ Rev.\ Lett.\  {\bf 81}, 1355 (1998)
  doi:10.1103/PhysRevLett.81.1355
  [hep-ph/9710415].


\bibitem{Liu:1993am} 
  J.~Liu and G.~Segre,
  Phys.\ Lett.\ B {\bf 338}, 259 (1994).
  doi:10.1016/0370-2693(94)91375-7


\bibitem{MarchRussell:1999ig} 
  J.~March-Russell, H.~Murayama and A.~Riotto,
  JHEP {\bf 9911}, 015 (1999)
  doi:10.1088/1126-6708/1999/11/015
  [hep-ph/9908396].


\bibitem{Iocco:2008va} 
  F.~Iocco, G.~Mangano, G.~Miele, O.~Pisanti and P.~D.~Serpico,
  Phys.\ Rept.\  {\bf 472}, 1 (2009)
  doi:10.1016/j.physrep.2009.02.002
  [arXiv:0809.0631 [astro-ph]].

\bibitem{Cyburt:2015mya} 
  R.~H.~Cyburt, B.~D.~Fields, K.~A.~Olive and T.~H.~Yeh,
  Rev.\ Mod.\ Phys.\  {\bf 88}, 015004 (2016)
  doi:10.1103/RevModPhys.88.015004
  [arXiv:1505.01076 [astro-ph.CO]].


\bibitem{Mangano:2011ip} 
  G.~Mangano, G.~Miele, S.~Pastor, O.~Pisanti and S.~Sarikas,
  Phys.\ Lett.\ B {\bf 708}, 1 (2012)
  doi:10.1016/j.physletb.2012.01.015
  [arXiv:1110.4335 [hep-ph]].


\bibitem{Lunardini:2000fy} 
  C.~Lunardini and A.~Y.~Smirnov,
  Phys.\ Rev.\ D {\bf 64}, 073006 (2001)
  doi:10.1103/PhysRevD.64.073006
  [hep-ph/0012056].


\bibitem{Dolgov:2002ab} 
  A.~D.~Dolgov, S.~H.~Hansen, S.~Pastor, S.~T.~Petcov, G.~G.~Raffelt and D.~V.~Semikoz,
  Nucl.\ Phys.\ B {\bf 632}, 363 (2002)
  doi:10.1016/S0550-3213(02)00274-2
  [hep-ph/0201287].


\bibitem{Wong:2002fa} 
  Y.~Y.~Y.~Wong,
  Phys.\ Rev.\ D {\bf 66}, 025015 (2002)
  doi:10.1103/PhysRevD.66.025015
  [hep-ph/0203180].


\bibitem{Abazajian:2002qx} 
  K.~N.~Abazajian, J.~F.~Beacom and N.~F.~Bell,
  Phys.\ Rev.\ D {\bf 66}, 013008 (2002)
  doi:10.1103/PhysRevD.66.013008
  [astro-ph/0203442].


\bibitem{Mangano:2010ei} 
  G.~Mangano, G.~Miele, S.~Pastor, O.~Pisanti and S.~Sarikas,
  JCAP {\bf 1103}, 035 (2011)
  doi:10.1088/1475-7516/2011/03/035
  [arXiv:1011.0916 [astro-ph.CO]].

  
\bibitem{Castorina:2012md} 
  E.~Castorina, U.~Franca, M.~Lattanzi, J.~Lesgourgues, G.~Mangano, A.~Melchiorri and S.~Pastor,
  Phys.\ Rev.\ D {\bf 86}, 023517 (2012)
  doi:10.1103/PhysRevD.86.023517
  [arXiv:1204.2510 [astro-ph.CO]].


 
\bibitem{NOvA}
https://www-nova.fnal.gov/

 
\bibitem{Maki:1962mu} 
  Z.~Maki, M.~Nakagawa and S.~Sakata,
  Prog.\ Theor.\ Phys.\  {\bf 28}, 870 (1962).
  doi:10.1143/PTP.28.870
 
\bibitem{Pontecorvo:1957cp} 
  B.~Pontecorvo,
  Sov.\ Phys.\ JETP {\bf 6}, 429 (1957)
  [Zh.\ Eksp.\ Teor.\ Fiz.\  {\bf 33}, 549 (1957)].
 
\bibitem{Sigl:1992fn} 
  G.~Sigl and G.~Raffelt,
  Nucl.\ Phys.\ B {\bf 406}, 423 (1993).
  doi:10.1016/0550-3213(93)90175-O

\bibitem{Pantaleone:1992eq} 
  J.~T.~Pantaleone,
  Phys.\ Lett.\ B {\bf 287}, 128 (1992).
  doi:10.1016/0370-2693(92)91887-F

\bibitem{Blaschke:2016xxt} 
  D.~N.~Blaschke and V.~Cirigliano,
  arXiv:1605.09383 [hep-ph].


\bibitem{deSalas:2016ztq} 
  P.~F.~de Salas and S.~Pastor,
  arXiv:1606.06986 [hep-ph].


\bibitem{Bajc:1999he} 
  B.~Bajc and G.~Senjanovic,
  Phys.\ Lett.\ B {\bf 472}, 373 (2000)
  doi:10.1016/S0370-2693(99)01432-X
  [hep-ph/9907552].

\bibitem{newCMBcons} 
  G.~Barenboim, W.~ H.~Kinney and W.~I.~Park,
   work in progress.




\end{thebibliography}
\end{document}